\documentclass[12pt]{iopart}

\newif\ifpdf
    \ifx\pdfoutput\undefined
    \pdffalse    
\else
    \pdfoutput=1 
    \pdftrue
\fi
\ifpdf
    \usepackage[pdftex]{graphicx}
\else
    \usepackage{graphicx}
\fi

\usepackage[english]{babel}
\usepackage{graphicx}
\usepackage{amsfonts}
\newcommand{\Sign}{\mathrm{Sign}}
\begin{document}
\ifpdf
    \DeclareGraphicsExtensions{.pdf}
\else
    \DeclareGraphicsExtensions{.eps}
\fi
\jl{6}

\title[WKB states of a dust shell]%
{WKB metastable quantum states of a de~Sitter--Rei\ss{}ner-Nordstr\"{o}m dust shell}

\author{Stefano Ansoldi\footnote[1]{Email address: ansoldi@trieste.infn.it}}

\address{Dipartimento di Fisica Teorica
dell'Universit\`a degli Studi di Trieste
and
I.N.F.N. Sezione di Trieste,
Strada Costiera 11, I-34014 Miramare (TS), Italy}

\begin{abstract}
We study the dynamics of a spherically symmetric dust shell
separating two spacetime domains, the \textit{interior} one
being a part of the de~Sitter spacetime and the exterior one
having the \textit{extremal} \textit{Rei\ss{}ner-Nordstr\"{o}m}
geometry. Extending the ideas of previous works on the subject,
we show that the it is possible to determine the
(metastable) WKB quantum states of this gravitational system.
\end{abstract}


\submitted

\maketitle

\section{\label{sec:int}Introduction}

The dynamics of relativistic thin shells is a recurrent topic in the
literature about the classical theory of gravitating systems and
the still ongoing attempts to obtain a coherent description of their quantum
behaviour. Certainly, a good reason to make this system a preferred one
for a lot of models is the clear, synthetic description
of its dynamics in terms of Israel's junction conditions
\cite{1966NuCiB..44..1.....I,1967NuCiB..48..463...I} (the null case,
considered in detail in the seminal paper
of C. Barrabes and W. Israel \cite{1991PhReD..43..1129..B},
is also interesting for null-like surfaces \cite{2000ReMaPh.46..399...J},
i.e. light-like matter shells \cite{2002PhReD..65..064036J}, for which an
Hamiltonian treatment is given in \cite{1998PhReD..57..2279..L}
generalizing the approach described in \cite{1994PhReD..50..3961..K}).
Using this formalism\footnote{But see also \cite{1970PhReLe.25..1771..G} and references
therein for a complementary approach which also tackles the issue of stability.},
which has an intuitive geometric meaning,
many relevant aspects of gravitation have been brought into light.

Gravitating shells have indeed been considered
as natural models for different astrophysical problems:
the description
of variable cosmic objects \cite{1997AsJ....482.963...N}
and of specific aspects
(like ejection \cite{1996PhLeA..214.227...N} or
crossing of layers \cite{1995JMaPh..36..3632..F},
critical phenomena \cite{2001BrJPh..31..188...W},
perturbations \cite{2001PhReD..6402024012M}
and back-reaction \cite{1999ClQuGr.16..131...A})
in gravitational collapse
\cite{1967NuCiA..LI..744...d,1968CzJPh..18..435...K,1970NuCiB..67..136...C}
are only a few examples.

Moreover, at larger scales, specific configurations of shells
have also been considered to construct cosmological models
\cite{2001GeReGr.33..531...M} (even with hierarchical (fractal)
structure \cite{1992AsJ....388.1.....R}), to analyze phase transitions
in the early universe \cite{1987PhReD..36..2919..B} or to describe cosmological
voids \cite{2000ClQuGr.17..2719..D}; semiclassical models
have tackled the problem of avoiding the initial singularity of
the Big-Bang scenario by quantum tunnelling
\cite{1990NuPhB..339.417...F,1991PhSc...T36.237...G,1997ClQuGr.14..2179..M}.

As a matter of fact quantum semiclassical models have conveniently been employed
as useful simple examples to better understand possible properties and
modifications to the spacetime structure at scales at which quantum effects
should give significant contributions to gravitational physics. Apart
from the quantization of the gravitating shell itself
(as in \cite{1997PhLeB..400.12....D,1988NuPhB..212.415...B}),
considered also in the context of gravitational
collapse \cite{1992PhReD..46..5439..H},
models have been proposed to study quantum properties of black
holes \cite{1990PhLeB..241.194...B,2002IJMPA..17..979...B,1996IJMPD..5...679...B}
and their formation process \cite{1996PhReD..53..4356..N} as well as to analyze
wormhole spacetimes \cite{1989NuPhB..328.203...V,1991PhReD..43..402...V,%
1995PhReD..52..6846..H,1995AmInPh...........V} and
the quantum stabilization of their instability
\cite{1995PhReD..52..7318..P,1990PhLeB..242.24....V},
targeting the fuzzy properties of spacetime foam \cite{1994PhReD..49..5199..R}
and Planck scale physics \cite{1994PhReD..49..3963..V}.

Other problems of fundamental nature in quantum gravity have received
attention through the study of shell dynamics:
as an exemplificative list, we mention here
Hawking radiation
\cite{1995NuPhB..433.403...K,1996IJMPD..5...679...B,1997PhReD..55..2139..B}
the horizon problem in wormhole spacetimes \cite{1993PhReLe.70..2665..H},
the time problem in canonical relativity \cite{2000PhReD..6204044025H},
the problem of localization of gravitational energy \cite{1990ClQuGr.7...787...K},
the thermodynamics of self-gravitating systems
\cite{1996PhReD..53..7062..M,1999PhReD..6012124018A}
and the possibility of connecting compact with non-compact dimensions
\cite{1991GeReGr.23..1415..G}.\\
Many of the above discussions have been performed under the simplifying assumption
of spherical symmetry: this is especially useful in the quantum treatment,
because the \textit{minisuperspace approximation} greatly reduces the complexity
of the mathematical treatment.\\
But, at least at the classical level,
studies have also been performed for cylindrical models (see e.g.
\cite{1994ClQuGr.11..167...G,2000GeReGr.32..2189..P,2000PhReD..6212124001P}).

With the development of the models shortly cited above, particularly those
involved with the quantization of the system, many subtleties emerged as
byproducts of corresponding difficulties already encountered in tentative
approaches to Quantum Gravity and mainly related to the reparametrization
invariance of the theory. Since the junction conditions, essentially, are
a first integral of the
equations of motion of the shell, many authors revolved their attention to
the derivation of these equations starting from an action principle.
A consistent Lagrangian/Hamiltonian formalism has been developed
\cite{1997PhReD..56..4706..H,1997PhReD..56..7674..F,%
1998PhReD..57..914...H,1999JMaPh..40..318...H} (also reduced by spherical symmetry
\cite{1998PhReD..57..936...H}),
and the relevant degrees of freedom of the system \cite{1998AcPhPoA29..1001..K}
discussed together with a variational principle, which is also the subject of
\cite{2001JMaPh..42..2590..G,2002PhReD..6502024028M} (interesting
considerations can also be found in \cite{1998PhReD..58..084005H}).

Recently even more interest in the thin shell formalism is coming thanks to the
development of brane world scenarios, where our universe is seen as a four
dimensional brane embedded in a five dimensional space
\cite{2000PhLeB..482.183...B,2000EuLe...49..396...G}. This configuration
can be given a wormhole interpretation \cite{2000PhReD..6206067502A}
and has also been analyzed from the point of view of energy conditions
\cite{2000NuPhB..584.415...B} (\textit{not}) satisfied in the higher
dimensional background.

In these and other studies, different cases of junctions between
spacetimes have been considered: for example between
anti de~Sitter and anti de~Sitter
    \cite{2000PhReD..6206067502A},
Friedmann-Robertson-Walker and Friedmann-Robertson-Walker
    \cite{1995PhReD..52..6846..H},
Minkowski and Minkowski
    \cite{1990PhLeB..242.24....V,1994PhReD..49..5199..R},
Schwarzschild and Schwarzschild
    \cite{1995PhReD..52..7318..P,1995JMaPh..36..3632..F,1997AsJ....482.963...N,1998PhReD..57..2279..L},
Rei\ss{}ner-Nordstr\"{o}m and Rei\ss{}ner-Nordstr\"{o}m
    \cite{1967NuCiA..LI..744...d,1991PhReD..43..402...V},
de~Sitter and Rei\ss{}ner-Nordstr\"{o}m
    \cite{2000NuPhB..584.415...B},
de~Sitter and Schwarzschild
    \cite{1987PhReD..35..1747..B,1991PhSc...T36.237...G,1999ClQuGr.16..3315..G},
de~Sitter and Schwarzschild-de~Sitter
    \cite{1992PrThPh.88..1097..Y},
de~Sitter and Vaidya
    \cite{1997ClQuGr.14..2179..M},
Friedmann-like and Rei\ss{}ner-Nordstr\"{o}m
    \cite{1994PhReD..49..2801..B},
Minkowski and Friedmann
    \cite{2000ClQuGr.17..2719..D},
Minkowski and Rei\ss{}ner-Nordstr\"{o}m
    \cite{1967NuCiA..LI..744...d,1997PhReD..55..2139..B,1998PhReD..57..4812..Z},
Minkowski and Schwarzschild
    \cite{1990PhLeB..241.194...B,1996PhReD..53..4356..N,1997PhLeB..400.12....D,1999ClQuGr.16..131...A,2000PhReD..6204044025H,2002IJMPA..17..979...B},
Minkowski and Vaidya
    \cite{2001BrJPh..31..188...W},
Schwarzschild and Rei\ss{}ner-Nordstr\"{o}m
    \cite{1998PhReD..57..4812..Z},
Schwarzschild and Schwarzschild-anti de~Sitter
    \cite{1991GeReGr.23..1415..G},
Schwarzschild and Vaidya
    \cite{1999PhReD..6012124018A},
Tolman and Friedman
    \cite{1992AsJ....388.1.....R},
Lema\^{\i}tre-Tolman-Bondi and Lema\^{\i}tre-Tolman-Bondi
    \cite{2001GeReGr.33..531...M}.

In this paper we are also going to use a general relativistic shell
to analyze, even if only at the semiclassical level, the problem of
quantization of a gravitational system. We will restrict ourselves,
as it has been done in many of the papers cited above, to the spherically
symmetric case and we will study the semiclassical quantum dynamics in the
case in which the shell separates an interior spacetime of the de~Sitter geometry,
from an exterior of the extremal Rei\ss{}ner-Nordstr\"{o}m type. An
observer crossing the shell will naively see some non-vanishing
vacuum energy density to be \textit{converted} into physical properties like
charge and mass. From the classical dynamics there are no restrictions
on the values of the physical parameters characterizing the geometry of
spacetime. But, starting from a Hamiltonian description of the shell dynamics,
we will try to analyze its quantum behaviour. Lacking a full theory of
quantum gravity, which would of course be the natural setting for this
kind of problem, we will tackle it only at the semiclassical
level: under this word, we will understand that the action for the shell
is given as an integer multiple of the quantum, $\hbar$. We will see that
this condition results in a constraint on the parameters for the
interior and exterior geometries. This is hardly surprising: indeed
a full quantum theory of gravity, would have the task of determining
the probability amplitude for a given configuration of the three-geometries
taught as points in superspace; in our quantum minisuperspace approach, the
only free parameters remain the constants (de~Sitter cosmological horizon,
charge and mass) fixing the interior and exterior metrics, and is thus
as a relation among them that the semiclassical quantization conditions
realizes itself.

With the above ideas in mind the paper is organized as follows.\\
In \sref{sec:pre} we will set up our model by giving all
relevant definitions; we will also recall some well known results
adapted to our special case, to fix notations and conventions,
and will present all relevant dynamical quantities for the computations
that follows. The Bohr--Sommerfeld quantization condition is
also recalled. Then, in \sref{sec:cladyn} the classical dynamics
of the system is sketched and the associated spacetime structure
discussed, with particular emphasis on the bounded
trajectories. This prepares the ground for
\sref{sec:claacteva}, where the classical action is numerically
evaluated for bounded trajectories. This result is then used in
\sref{sec:WKBquadyn} to show how the Bohr--Sommerfeld
quantization condition characterizes the properties of the
semiclassical quantum system. After a preliminary rough
estimate (subsection \ref{sec:WKBquadynpreext}), we present
the semi-classical results for the quantum levels of the shell and
the corresponding internal/external geometries and approximate the
results with a properly chosen analytic (polynomial) expression.
Discussion about the results and possible refinements of the model follow
in \sref{sec:dis}.
\\
Five short appendices are devoted to a more detailed
analysis of some technical points. The turning points of the classical motion are
discussed in \ref{app:potcritic}. The issue about the stability
of the classical solution against single particle decay is studied in
\ref{app:staiss}. \ref{app:sigintsigout}
shows that the bounded trajectories are not affected by change of direction of the
normal to the shell trajectory.
The characterization of the singularity that appears in the integral
for the computation of the classical action as an integrable one is done in
\ref{app:sinintpat} and the determination of the leading terms
in the integrand of the same computation is the topic of \ref{app:powexp}.

\section{\label{sec:pre}Preliminaries}

In this section we define the system, motivate
the settings under which we study its classical
and semiclassical dynamics and recall some useful results
and definitions.

Let us thus start with the geometrodynamical framework, by considering
two spacetime domains joined along a spherically symmetric timelike shell.
We assume, for the region we shall call the \textit{interior},
a geometry of the de~Sitter type
\cite{1917PrKoNeA19..1217..D,1917PrKoNeA20..229...D,1970WHFranCGr........M}
(we denote with $H$ the cosmological horizon),
so that the metric in static coordinates is:
\begin{eqnarray}
    g _{\mathrm{in}} ^{\mu \nu}
    =
    \mathrm{diag}
    \left(
        f _{\mathrm{in}} (r)
        ,
        f _{\mathrm{in}} ^{-1} (r)
        ,
        r^{2}
        ,
        r^{2} \sin \theta
    \right)
    \label{eq:metin}
    \\
    f _{\mathrm{in}} (r)
    =
    1 - \frac{r ^{2}}{H ^{2}}
    \nonumber
    .
\end{eqnarray}
For the \textit{exterior} region we choose a spacetime of the
Rei\ss{}ner-Nordstr\"{o}m type
\cite{1916AnPhGe.50..106...R,1918PrKoNeA20..1238..N,1970WHFranCGr........M},
with metric given by
\begin{eqnarray}
    g _{\mathrm{out}} ^{\mu \nu}
    =
    \mathrm{diag}
    \left(
        f _{\mathrm{out}} (r)
        ,
        f _{\mathrm{out}} ^{-1} (r)
        ,
        r^{2}
        ,
        r^{2} \sin \theta
    \right)
    \label{eq:metout}
    \\
    f _{\mathrm{out}} (r)
    =
    1 - \frac{2 M}{r} + \frac{Q ^{2}}{r ^{2}}
    \nonumber
    ,
\end{eqnarray}
$M$ being the Schwarzschild mass and $Q$ the electric
charge.
Furthermore we join the two regions along the timelike trajectory
of a spherical dust shell of constant total mass-energy $m$.

As is well known \cite{1966NuCiB..44..1.....I,1967NuCiB..48..463...I}
the dynamics of the compound gravitational system is
encoded in Israel's junction conditions: they match the jump in the
extrinsic curvature due to the different spacetime geometries on the two
sides of the shell surface and the (singular) stress-energy tensor
of the shell itself. Under the simplifying assumption of
spherical symmetry considered here, it is possible to reduce them
to the single scalar equation
\cite{1997ClQuGr.14..2727..A,1994DeThPh.1...169...A}
\begin{equation}
    \left[
        \sigma \beta
    \right]
    =
    \frac{m}{R}
    ,
\label{eq:isrjuncon}
\end{equation}
where as customary we use square brackets as a shorthand for the jump of the
enclosed quantity in the passage from the ``in'' to the ``out'' domain across the
shell\footnote{To avoid any possible
confusion, in what follows we are going to use
square brackets \textit{only} with this meaning, according to
the following definition.},
i.e.
\[
    \left[ X \right] := X _{\mathrm{in}} - X _{\mathrm{out}}
    ,
\]
and
\begin{eqnarray}
    \left( \sigma \beta \right) _{\mathrm{in}}
    :=
    \sigma _{\mathrm{in}} \beta _{\mathrm{in}}
    =
    \sigma _{\mathrm{in}} \sqrt{\dot{R} ^{2} + 1 - \frac{R ^{2}}{H ^{2}}}
    \nonumber \\
    \left( \sigma \beta \right) _{\mathrm{out}}
    :=
    \sigma _{\mathrm{out}} \beta _{\mathrm{out}}
    =
    \sigma _{\mathrm{out}} \sqrt{\dot{R} ^{2} + 1 - \frac{2 M}{R} + \frac{Q ^{2}}{R ^{2}}}
    .
    \nonumber
\end{eqnarray}
In the above expressions $R = R ( \tau )$ is the shell radius expressed
as a function of
the proper time $\tau$ of an observer co-moving with the shell and
we denote with an over-dot the (total) derivative with respect to $\tau$.

From the general theory of shell dynamics, we know that the signs
of the radicals \textit{do} matter, being related both to the
side of the maximally extended diagram for the spacetime manifold, which
is crossed by the trajectory of the shell \cite{1987PhReD..35..1747..B},
and to the direction of the
outward (i.e toward increasing radius)
pointing normal to the shell surface
\cite{1966NuCiB..44..1.....I,1967NuCiB..48..463...I}.
This is the reason why they are
denoted explicitly by $\sigma _{\mathrm{in}/\mathrm{out}}$.
Their values can be analytically determined thanks to the results
\cite{1997ClQuGr.14..2727..A,1994DeThPh.1...169...A}
\begin{eqnarray}
    \sigma _{\mathrm{in}}
    =
    \sigma _{\mathrm{in}} (R)
    =
    -
    \Sign
    \left\{
        m
        \left(
            \frac{R ^{4}}{H ^{2}}
            -
            2 M R
            +
            Q ^{2} - m ^{2}
        \right)
    \right\}
    \label{eq:sigsigin}
    \\
    \sigma _{\mathrm{out}}
    =
    \sigma _{\mathrm{out}} (R)
    =
    -
    \Sign
    \left\{
        m
        \left(
            \frac{R ^{4}}{H ^{2}}
            -
            2 M R
            +
            Q ^{2} + m ^{2}
        \right)
    \right\}
    ,
    \label{eq:sigsigout}
\end{eqnarray}
which can be obtained by properly squaring the junction condition
(\ref{eq:isrjuncon}).

Following the notation of reference
\cite{1997ClQuGr.14..2727..A} we know that the
junction condition (\ref{eq:isrjuncon}) can be derived as the Superhamiltonian
constraint for the system, where the corresponding Superhamiltonian is then
nothing but
\begin{equation}
    {\mathcal{H}}
    =
    P \dot{R} - {\mathcal{L}}
    =
    R
    \left[ \sigma \beta \right]
    -
    m
    ,
    \label{eq:ham}
\end{equation}
${\mathcal{L}}$ being the Lagrangian density
\begin{equation}
    {\mathcal{L}}
    =
    m
    -
    R
    \left\{
        \left[
            \sigma \beta
            -
            \frac{1}{2}
            \dot{R}
            \ln
            \left|
                \frac{\sigma \beta + \dot{R}}{\sigma \beta - \dot{R}}
            \right|
        \right]
    \right\}
    \label{eq:lag}
\end{equation}
and $P$ being the conjugate momentum to the canonical variable
$R$:
\begin{equation}
    P
    =
    \frac{\partial {\mathcal{L}}}{\partial \dot{R}}
    =
    -
    \frac{R}{2}
    \left\{
        \left[
            \ln
            \left|
                \frac{\sigma \beta + \dot{R}}{\sigma \beta - \dot{R}}
            \right|
        \right]
    \right\}
    .
    \label{eq:momvel}
\end{equation}
The dynamics of the system can be studied with the help of an effective
equation of motion, which is useful in removing the square roots
in (\ref{eq:isrjuncon}) and can be put in the form of a classical
one-dimensional dynamical problem, the motion of an effective particle
of unitary mass and vanishing total energy
\cite{1987PhReD..35..1747..B,1989PhReD..40..2511..A}
\begin{equation}
    \dot{R} ^{2} + V (R) = 0
\label{eq:effequ}
\end{equation}
in a potential given by
\begin{eqnarray}
    \fl
    V (R)
    =
    -
    \frac{
          R ^{8}
          -
          4 H^{2} M R ^{5}
          +
          2 H ^{2} (m ^{2} + Q ^{2}) R ^{4}
         }
         {4 m ^{2} H ^{2} R ^{2}}
    \nonumber \\
    -
    \frac{
          4 H ^{4} (M ^{2} - m ^{2}) R ^{2}
          +
          4 H ^{4} M (m ^{2} - Q ^{2}) R
          +
          H ^{4} (m ^{2} - Q ^{2}) ^{2}
         }
         {4 m ^{2} H ^{2} R ^{2}}
    .
    \label{eq:claequpot}
\end{eqnarray}
To evaluate the classical action along a classically allowed
trajectory, we need an expression for the effective momentum
$P$ evaluated along the same trajectory. Thus we have to substitute
for the $\dot{R}$ dependence in (\ref{eq:momvel}) and using
in this procedure relation (\ref{eq:effequ}) we get
\begin{eqnarray}
    \fl
    P (R)
    =
    -
    R
    \tanh ^{-1}
    \left\{
        \left(
            \frac{2 m H ^{2} R \sqrt{- V (R)}}
                 {
                  -
                  R ^{4}
                  +
                  2 H ^{2} R ^{2}
                  -
                  2 H^{2} M R
                  +
                  H ^{2} \left( Q ^{2} - m ^{2} \right)
                 }
        \right) ^{s (r)}
    \right\}
    ,
    \label{eq:genmom}
    \\
    \fl
    \mathrm{with}
    \quad
    s (r)
    =
    \Sign
    \left\{
        \left( 1 - \frac{R ^{2}}{H ^{2}} \right)
        \left( 1 - \frac{2 M}{R} + \frac{Q ^{2}}{R ^{2}} \right)
    \right\}
    .
    \nonumber
\end{eqnarray}
We can now use the above result to compute the action
along a classically allowed trajectory, having turning points at
$R _{1}$ and $R _{2}$, since
\begin{equation}
    S _{\mathrm{classical}}
    =
    2
    \int _{R _{1}} ^{R _{2}}
        P (R) \rmd R
    ,
\label{eq:claacteva}
\end{equation}
and then implement a semiclassical quantization scheme \textit{a la}
Bohr--Sommerfeld, by considering allowed quantum states to
have the action as an integer multiple of the elementary quantum
\cite{1982SpVe...1...133...M}
$l _{\mathrm{P}} ^{2} \equiv \hbar \equiv 1$:
\begin{equation}
    S _{\mathrm{classical}}
    =
    n
    ,
    \quad
    n = 1 , 2 , \dots .
    \label{eq:bohsomquaconint}
\end{equation}
To successfully complete this task we need in first place an analysis of
the allowed bounded classical trajectories, which we will perform
in the next section.

Before embarking this program, let us shortly
comment about the semiclassical quantization procedure outlined above. There
are indeed many different approaches for the quantization of gravitational
systems, and it is not often clear which should be the preferred one. Moreover,
deep ideas have already been discussed to a great extent in the literature
cited above. It is not the goal of this paper to address this fundamental
problem, but we think it is important to give a short account about the reliability
of the results that will be derived in what follows. In particular a formalism
using expression \eref{eq:genmom}, but evaluated along a classically forbidden
trajectory, has already been successfully used in \cite{1996PhEs...9...556...A} and
\cite{1997ClQuGr.14..2727..A} to reproduce some well known results about
vacuum decay and the influence of gravity on it, already studied in the
seminal papers by Coleman and de~Luccia \cite{1980PhReD..21..3305..C}
and by Parke \cite{1983PhLeB..121.313...P}.
Moreover, as already noted by Sommerfeld in the days of the
early development of Quantum Mechanics \cite{1982SpVe...1...133...M},
the quantization condition \eref{eq:bohsomquaconint} is
``\textit{particularly valuable, for it could be applied both to relativistic
and non-relativistic systems\/}''. Thus, we think that the above considerations
justify our tentative approach, in which the semiclassical quantization
condition is applied, through an already tested procedure, to a
classically well-know gravitational system.

\section{Classical Dynamics}
\label{sec:cladyn}

\begin{figure}
    \begin{center}
        \fbox{\includegraphics[width=5cm]{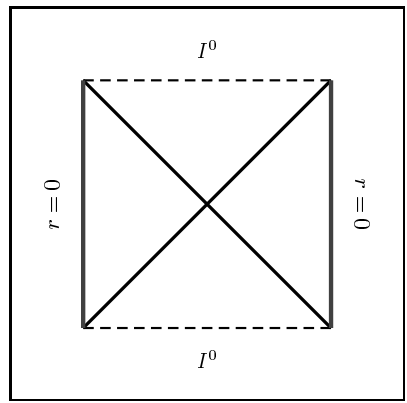}}
    \end{center}
    \caption{Maximally extended Penrose diagram of the de~Sitter spacetime.}
    \label{fig:deSPendia}
\end{figure}

\begin{figure}
    \begin{center}
        \fbox{\includegraphics[width=5cm]{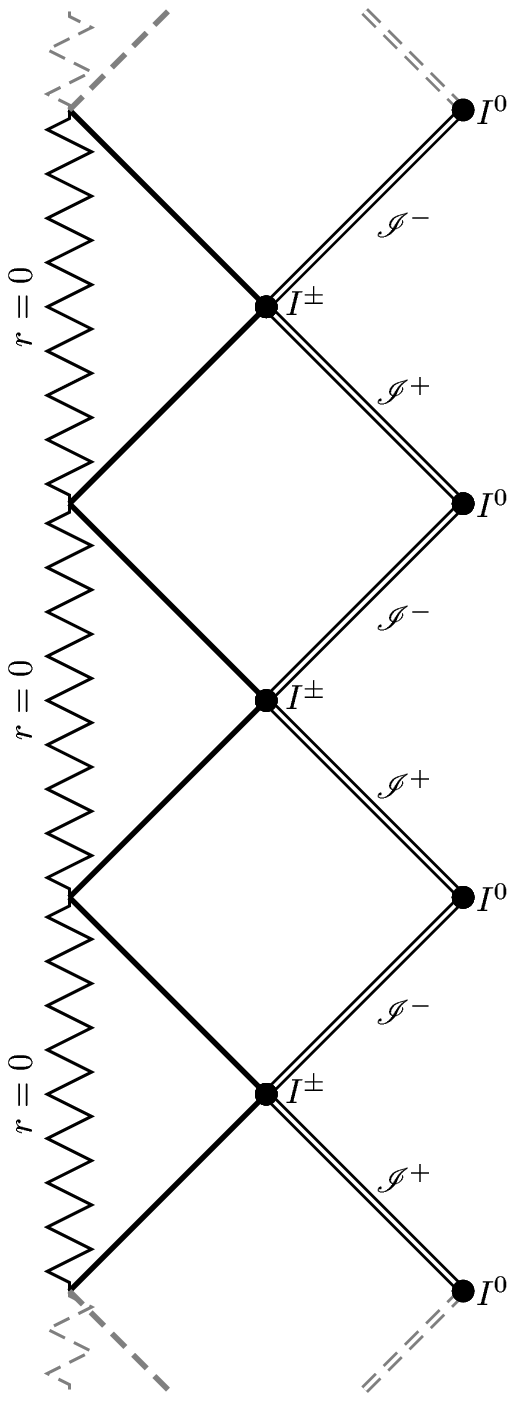}}
    \end{center}
    \caption{Maximally extended Penrose diagram of the extremal
    Rei\ss{}ner-Nordstr\"{o}m spacetime.}
    \label{fig:R-NPendia}
\end{figure}

The results presented above are valid for arbitrary values of the four
parameters entering the problem, namely the mass $M$ and the charge $Q$ of
the external Rei\ss{}ner-Nordstr\"o{}m spacetime,
the de~Sitter radius $H$ of the interior
geometry and the total mass-energy of the dust shell, $m$. We now
specialize them to a more particular setting (which has the
advantage of removing the second line in expression \eref{eq:claequpot}
for the effective potential):
\begin{enumerate}
    \item we take the external Rei\ss{}ner-Nordstr\"{o}m
    spacetime to be \textit{extremal}, i.e. with $| Q | = M$;
    \item we assume that the total mass energy of the shell is
    $m = | Q |$.
\end{enumerate}
The Penrose diagrams for the full de~Sitter and extremal
Rei\ss{}ner-Nordstr\"{o}m \cite{1966PhLe...21..423...C}
spacetimes are shown in figures
\ref{fig:deSPendia} and \ref{fig:R-NPendia} respectively.

Before studying the possible shell trajectories in the two geometries
to identify the bounded ones, which we are interested in, we take full advantage
of the parameter reduction implicit in the assumptions above, by passing
to adimensional variables: this will be more convenient also for the
subsequent numerical treatment. We thus choose
to parametrize all the variables and constants
in terms of the de~Sitter cosmological horizon $H$ by setting
\begin{equation}
    x = \frac{R}{H}
    \quad , \quad
    t = \frac{\tau}{H}
    \quad , \quad
    \Theta = \frac{Q}{H}
    \quad , \quad
    | \Theta | = \frac{M}{H} = \frac{m}{H}
    .
\label{eq:admdef}
\end{equation}
Then the quantities which are functions of $R$, become functions of $x$,
all retaining their numerical values but $P$ and $S$, which are rescaled by
$H$ and $H ^{2}$ respectively:
\begin{equation}
    \bar{P} ( x ; \Theta ) = \frac{P ( R ; H , Q )}{H}
    \quad , \quad
    \bar{S} ( \Theta ) = \frac{S ( H , Q )}{H ^{2}}
    .
\label{eq:admPS}
\end{equation}
We thus have for the quantities evaluated along a classical trajectory,
which are relevant in the study
of the classical dynamics\footnote{We denote with an overbar quantities,
let us say $g$,
which are function of the rescaled radial coordinate $x$, although in
many cases we have $\bar{g} (x) = g (R)$. For the sake of
precision, note that
$f _{\mathrm{in}} (R) = \bar{f} _{\mathrm{in}} (x)$,
$f _{\mathrm{out}} (R) = \bar{f} _{\mathrm{out}} (x)$,
$\sigma _{\mathrm{in}} (R) = \bar{\sigma} _{\mathrm{in}} (x)$,
$\sigma _{\mathrm{out}} (R) = \bar{\sigma} _{\mathrm{out}} (x)$,
and
$V (R) = \bar{V} (x)$.}
\begin{eqnarray}
    \fl
    \bar{f} _{\mathrm{in}} (x)
    =
    1 - x ^{2}
    \label{eq:admmetin}
    \\
    \fl
    \bar{f} _{\mathrm{out}} (x)
    =
    1 - \frac{2 | \Theta |}{x} + \frac{\Theta ^{2}}{x ^{2}}
    \label{eq:admmetout}
    \\
    \fl
    \bar{\sigma} _{\mathrm{in}} (x)
    =
    -
    \Sign \left( x ^{3} - 2 | \Theta | \right)
    \label{eq:admsigin}
    \\
    \fl
    \bar{\sigma} _{\mathrm{out}} (x)
    =
    -
    \Sign \left( x ^{4} - 2 | \Theta | x + 2 \Theta ^{2} \right)
    \label{eq:admsigout}
    \\
    \fl
    \bar{V} (x)
    =
    -
    \frac{x ^{2} \left( x ^{4} - 4 | \Theta | x + 4 \Theta ^{2} \right)}{4 \Theta ^{2}}
    \label{eq:admpot}
    \\
    \fl
    \bar{P} (x)
    =
    -
    x
    \tanh ^{-1}
    \left\{
        \left(
            \frac{x \sqrt{x ^{4} - 4 | \Theta | x + 4 \Theta ^{2}}}
                 {-x ^{3} + 2 x - 2 | \Theta |}
        \right) ^{
                  \Sign
                  \left\{
                    \left( 1 - x ^{2} \right)
                    \left( x  - | \Theta | \right) ^{2}
                  \right\}
                 }
    \right\}
    ;
    \label{eq:admmom}
\end{eqnarray}
of course all can be expressed as functions of the single adimensional
parameter $\Theta$.

\begin{figure}
    \begin{center}
        \fbox{\includegraphics[width=10cm]{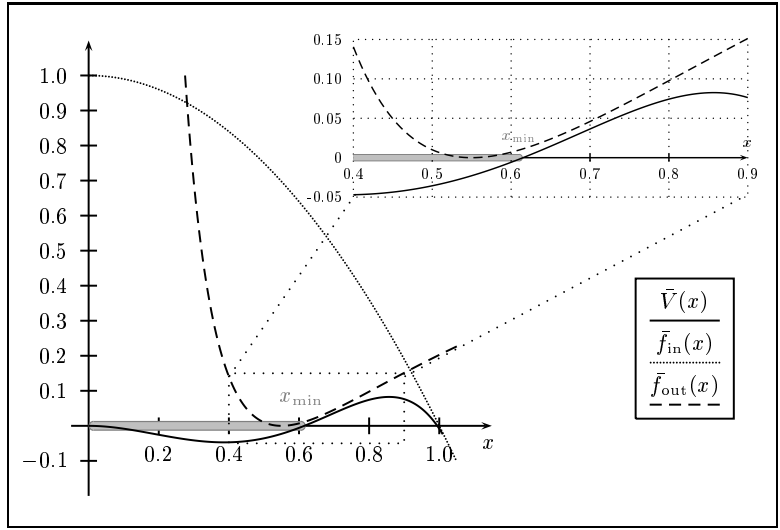}}
    \end{center}
    \caption{The graphical method to study the classical
    shell trajectories consists in plotting the potential
    $\bar{V} (x)$, together with the curves for the metric functions
    of the interior and of the exterior. The light gray path is a
    classically allowed bounded trajectory: as discussed in the text it
    crosses the horizon of the external geometry but no changes
    in the orientation of the normals occur along it.}
    \label{fig:clastu}
\end{figure}

\begin{figure}
    \begin{center}
        \fbox{\includegraphics[width=5cm]{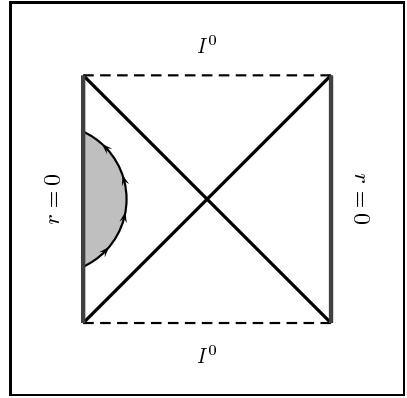}}
    \end{center}
    \caption{Penrose diagram of the de~Sitter interior geometry with
    the bubble trajectory. The interior domain is the shaded region
    in the diagram.}
    \label{fig:deSPendiatra}
\end{figure}

\begin{figure}
    \begin{center}
        \fbox{\includegraphics[width=5cm]{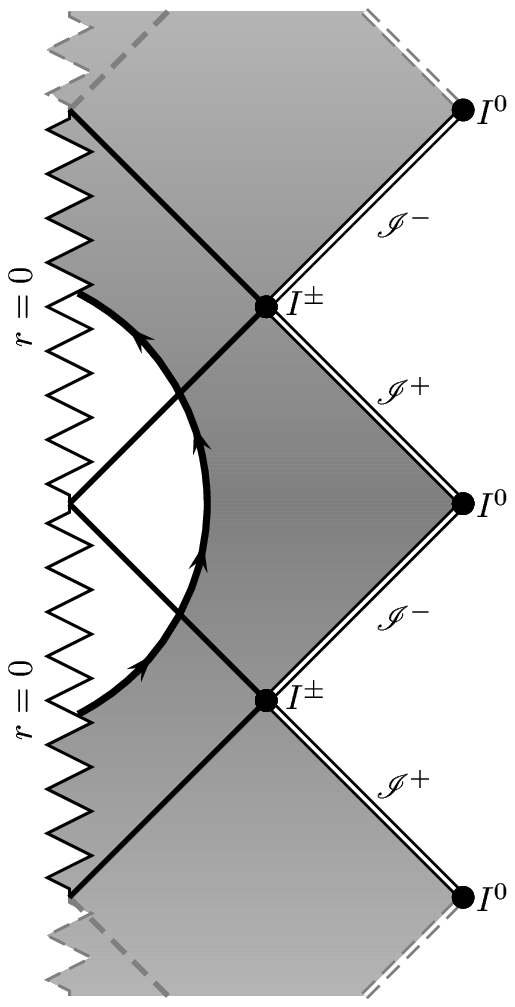}}
    \end{center}
    \caption{Penrose diagram of the Rei\ss{}ner-Nordstr\"{o}m exterior geometry with
    the bubble trajectory. The exterior domain is the shaded region in
    the diagram.}
    \label{fig:R-NPendiatra}
\end{figure}

Following \cite{1987PhReD..35..1747..B,1989PhReD..40..2511..A}
we can study the classical
dynamics in a compact way by means of a comprehensive graphical
method fully exploiting the handy relation \eref{eq:effequ}.
It consists in plotting the potential $\bar{V} (x)$ together with
the metric functions $\bar{f} _{\mathrm{in}}$, $\bar{f} _{\mathrm{out}}$. Then the
allowed trajectories with the corresponding turning points can be determined looking
at the segments of the $x$-axis (corresponding to zero energy), that are above
the graph of the potential. In this diagram the points where the metric functions vanish,
quickly help in determining if a classical path crosses the horizons of the
external/internal geometry. Moreover the rescaled values of
the radial coordinate $x$ for which $\sigma _{\mathrm{in/out}}$
changes sign are given, if they exist, by
the values at which the metric function plots are tangent to
the graph of the potential. This graphical information can be completed by
the following analytical results.
\begin{description}
    \item[Turning points of the potential]:\\
    from \eref{eq:admpot} we see that the potential for
    $\Theta \neq 0$ has a (double) zero at $x = 0$, so that it is regular
    at the origin, which is thus a trivial turning point of classical trajectories.
    Other turning points may, or may not, be present, depending on the value
    assumed by the parameter $\Theta$. Two cases are possible, as shown in
    \fref{fig:potcurves}.
    \begin{figure}
        \begin{center}
            \fbox{\includegraphics[width=6cm]{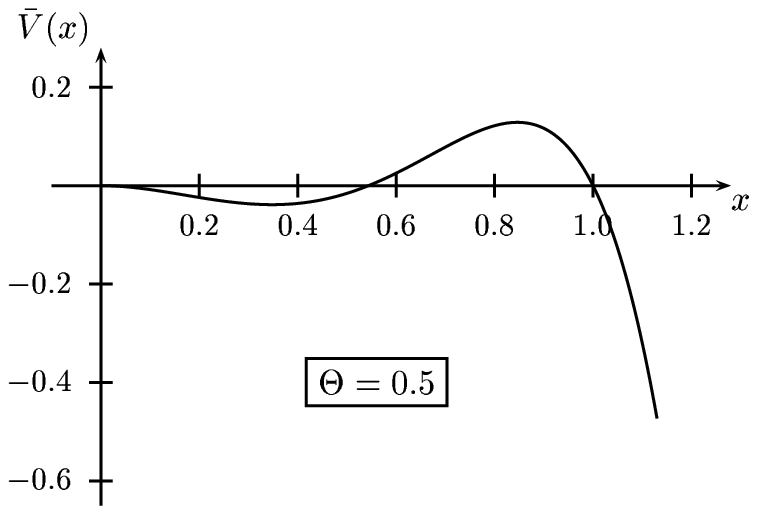}}
            \\[5mm]
            \fbox{\includegraphics[width=6cm]{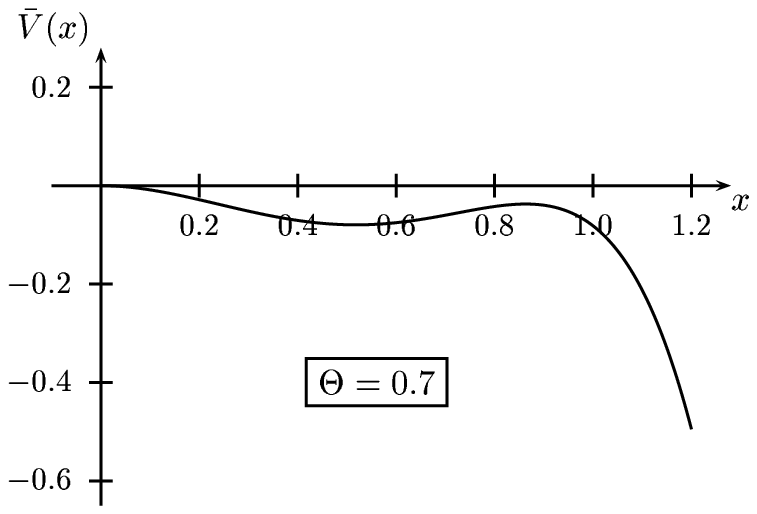}}
            \\[5mm]
            \fbox{\includegraphics[width=6cm]{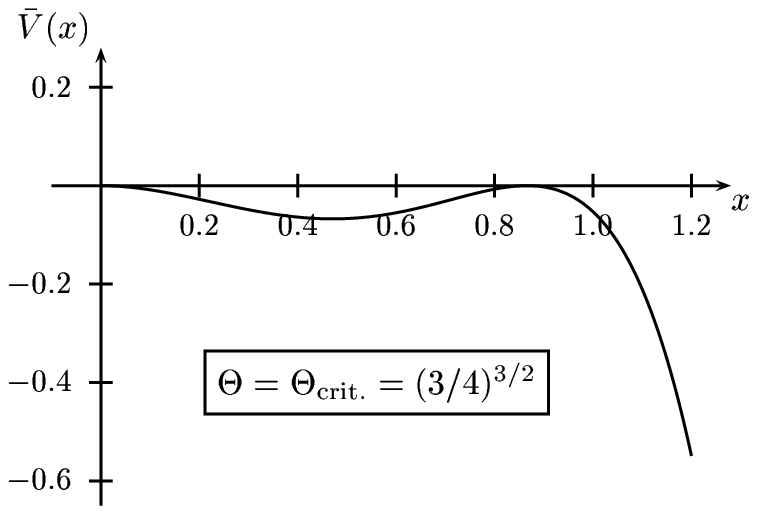}}
        \end{center}
        \caption{Graph of the potential $\bar{V} (x)$ for different values
        of the parameter $\Theta$. Depending on the value of $\Theta$, the
        classical trajectory can have two non-vanishing turning points, or
        no non-vanishing turning points, as shown in the first and second
        figure, respectively. The in between case, which occurs for
        $\Theta = \Theta _{\mathrm{crit.}} = (3/4) ^{3/2}$, is depicted in
        the third diagram.}
        \label{fig:potcurves}
    \end{figure}
    Either there can be no other turning points, so that only a so called ``bounce''
    classical trajectory exists (as is the case in \fref{fig:potcurves}
    for $\Theta = 0.7$), or there can be two more turning points so that in addition
    to the bounce trajectory there is also a bounded one (this is also shown in
    \fref{fig:potcurves} for $\Theta = 0.5$). As explicitly proved
    in \ref{app:potcritic}, the critical value for the
    parameter $\Theta$, $\Theta _{\mathrm{crit.}} = ( 3 / 4 ) ^{3/2}$
    gives the \textit{in between} case, when the potential is tangent
    to the $x$ axis (third plot, again in \fref{fig:potcurves}).
    We thus see that only for $0 < \Theta \leq \Theta _{\mathrm{crit.}}$
    there are two non-vanishing turning points $x _{\mathrm{min}}$,
    $x _{\mathrm{max}}$ (actually with $x _{\mathrm{min}} \equiv x _{\mathrm{max}}$
    if $\Theta = \Theta _{\mathrm{crit.}}$)
    and thus bounded solutions are allowed, the classical path
    being represented by the segment $[0 , x _{\mathrm{min}} ]$.
    We note that it is possible to find $x _{\mathrm{min}}$ and
    $x _{\mathrm{max}}$ in closed form solving the quartic equation that gives
    the non-vanishing solutions of $\bar{V} (x) = 0$, and
    this (not very enlightening) expressions are reported in \ref{app:potcritic}.
    \item[Horizon positions with respect to the classical path]:\\
    to correctly understand the spacetime geometry we also need the relative positions
    of the horizons with respect to the classical trajectories. This is also briefly
    discussed later on, but it is useful to report here a general result that
    can be deduced from \fref{fig:turpoiandhor}, where for
    $0 < \Theta \leq \Theta _{\mathrm{crit.}}$ the turning points, $0$,
    $x _{\mathrm{min}}$, $x _{\mathrm{max}}$ and the horizon of
    the exterior metric ($x (\Theta) = \Theta$) are plotted as functions of $\Theta$.
    \begin{figure}
        \begin{center}
            \fbox{\includegraphics[width=6cm]{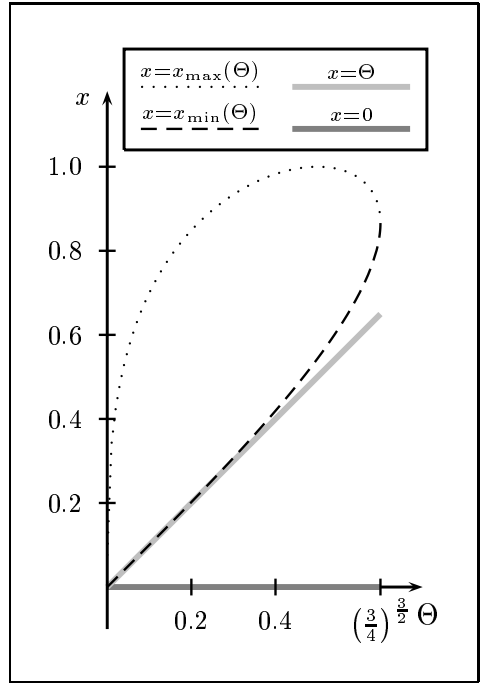}}
        \end{center}
        \caption{Graph of the non-negative roots ($x = 0$, $x = x _{\mathrm{min}} (\Theta)$
        and $x = x _{\mathrm{max}} (\Theta)$) of the potential $\bar{V} (x)$
        together with the horizon $x = \Theta$ of the ``out'' Rei\ss{}ner-Nordstr\"{o}m
        spacetime. Bounded trajectories are delimited by the $x = 0$ line and the
        $x = x _{\mathrm{min}} (\Theta)$ curve, so that they always cross the horizon
        of the Rei\ss{}ner-Nordstr\"{o}m spacetime, which the graph shows to be
        always smaller than $x = x _{\mathrm{min}} (\Theta)$.}
        \label{fig:turpoiandhor}
    \end{figure}
    It can be seen that the classical bounded trajectory, corresponding to the region
    between the $x$ axis and the $x _{\mathrm{min}}$ dashed curve, crosses
    for all values of $\Theta$ in the considered range the exterior
    horizon at $x = \Theta$. The horizon of the
    internal de~Sitter domain (which is not plotted and corresponds to the
    horizontal line $x \equiv 1$ in rescaled variables) is instead never crossed
    by a bounded trajectory\footnote{Nevertheless
    it equals $x _{\mathrm{max}}$ for $\Theta = 1 / 2$.
    This is an interesting limiting situation in the case of
    tunnelling across the potential barrier, which will be discussed elsewhere.}.
    \item[Asymptotic behaviour]:\\
    quite generally, from \eref{eq:admpot} we also see that
    \begin{equation}
        \lim _{x \to + \infty}
            \bar{V} (x)
        =
        - \infty
        .
    \label{eq:potliminf}
    \end{equation}
    \item[Regularity at the origin]:\\
        the first and second derivatives of the potential are vanishing
        at $x = 0$,
        $\rmd \bar{V} (x) / \rmd x = \rmd ^{2} \bar{V} (x) / \rmd x ^{2} = 0$,
        and the third derivative is positive,
        $\rmd ^{3} \bar{V} (x) / \rmd x ^{3} = 6 / | \Theta |$, so that $x = 0$ is a local
        maximum for $\bar{V} (x)$.
\end{description}
Thanks to the above properties, we can now perform the study of the classical
dynamics for bounded trajectories by restricting the parameter $\Theta$ to the
range
\begin{equation}
    0 < | \Theta | \leq \left( \frac{3}{4} \right) ^{3/2}
    ,
\label{eq:theranboutra}
\end{equation}
where we have the general situation shown in \fref{fig:clastu}.
The figure shows, for $\Theta = 0.55$, the potential $\bar{V} (x)$ together with
$\bar{f} _{\mathrm{in}} (x)$ and $\bar{f} _{\mathrm{out}} (x)$. The classically
allowed path is the thicker light gray segment on the $x$-axis.
Let us consider the dynamics in the de~Sitter spacetime: the shell expands
from vanishing radius up to a maximum (grayed path  in the figure),
which remains inside the de~Sitter
cosmological horizon, since, as can be seen, it is not crossed by the trajectory;
then the shell shrinks back to zero radius. The sign of $\sigma _{\mathrm{in}}$
does not change along the trajectory, since as we can see always from
\fref{fig:clastu}, there are no points on the trajectory
in which the plot of
$\bar{f} _{\mathrm{in}} (x)$ is tangent to $\bar{V} (x)$: moreover, as can be
easily verified, it is always positive, so that the trajectory
crosses the left part of the de~Sitter Penrose diagram. This is shown
in \fref{fig:deSPendiatra}, where the interior region is the shaded area,
since for $\sigma _{\mathrm{in}} = + 1$ the exterior normal is
pointing to the right.
In the same way we can perform the analysis in the Rei\ss{}ner-Nordstr\"{o}m
domain: we can see (in \fref{fig:clastu}, but also from the
above discussion about \fref{fig:turpoiandhor}) that the bounded
trajectory during the expansion from a vanishing radius,
as well as during the following collapse, crosses the horizon of
the exterior geometry. As before the sign of $\sigma _{\mathrm{out}}$
does not change along the trajectory (in the zoomed region of \fref{fig:clastu} we
can more clearly see that the metric function graph is \textit{not} tangent
to the potential) and thus $\sigma _{\mathrm{out}} = - 1$ always. If we
draw the associated Penrose diagram, the exterior region is
the shaded one in \fref{fig:R-NPendiatra}.
The results shown above rest on two hypotheses, that we implicitly
took for granted but which deserve a more detailed treatment.

The first one concerns the stability of the expanding and recollapsing
shell against single particle decay: if the shell were not stable at the
moment of time symmetry, then it would become thick and its trajectory
would not be approximated by a sharp line (as depicted in figures
\ref{fig:deSPendiatra} or \ref{fig:R-NPendiatra}).
It is possible to show, following the treatment of
\cite{1970PhReLe.25..1771..G} (please see \ref{app:staiss} for details),
that configurations stable against single particle decay of
charged and/or uncharged particles actually exist.

The second issue is about possible changes of signs in
$\sigma _{\mathrm{in}}$ or $\sigma _{\mathrm{out}}$, which would require
a different analysis with respect to the one performed above. We also devote
an appendix (\ref{app:sigintsigout}) to show that
bounded trajectories are not affected by changes of sign in
$\sigma _{\mathrm{in}}$ or $\sigma _{\mathrm{out}}$, so that the analysis
performed above in a particular case is indeed valid in general.

\begin{figure}
    \begin{center}
        \fbox{\includegraphics[width=5cm]{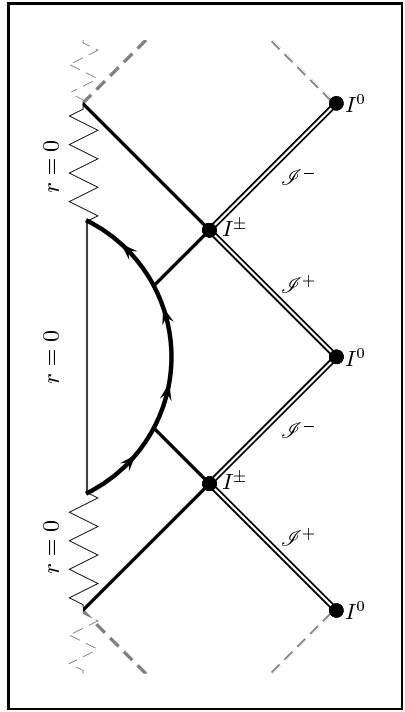}}
    \end{center}
    \caption{Penrose diagram of the de~Sitter interior and
    the Rei\ss{}ner-Nordstr\"{o}m exterior geometries joined along
    the bubble trajectory. It is obtained by joining the shaded
    regions of the two previous figures. The classical dynamics
    of this compound spacetime is described by Israel's junction
    conditions and is discussed in detail in the text.}
    \label{fig:deSR-NPendiatra}
\end{figure}

With these remarks in mind, the complete spacetime manifold can now be confidently
obtained by joining the interior with the exterior along the shell trajectory,
i.e. joining the two shaded regions in \fref{fig:deSPendiatra} and
\fref{fig:R-NPendiatra} to get the final result shown in \fref{fig:deSR-NPendiatra}.
An observer inside the shell detects a non-vanishing
cosmological constant. He lives as an observer inside a cosmological horizon.
But as soon as he crosses the shell trajectory, he experiences
a completely different situations: the cosmological constant suddenly vanishes
and he can now detect a non-vanishing electric and gravitational field,
as if outside a body with mass $M$ and charge $Q$ (with $M = | Q |$, because
of our simplifying assumptions).
We are now interested in studying the properties of this gravitational configuration
when the system can be considered to be in a semi-classical quantum regime.
For this we need an evaluation of the classical action along the classical trajectory
of the shell.

\section{\label{sec:claacteva}Numerical evaluation of the classical action}

As already anticipated in \eref{eq:claacteva} we will evaluate
the classical action as the integral of the classical momentum
along a classically allowed trajectory. In our case, remembering
the naming conventions about the zeros of $\bar{V} (x)$, this implies
that the relevant turning points for the bounded trajectory in rescaled
coordinates are
$R _{1} / H \equiv x _{1} = 0$ and
$R _{2} / H \equiv x _{2} = x _{\mathrm{min}}$, so that
\begin{eqnarray}
    \fl
    \bar{S}
    =
    \frac{S}{H ^{2}}
    =
    2
    \int _{0} ^{x _{\mathrm{min}}}
        \bar{P} (x) \rmd x
    \nonumber \\
    =
    - 2
    \int _{0} ^{x _{\mathrm{min}}}
        x
        \tanh ^{-1}
        \left\{
            \left(
                \frac{x \sqrt{x ^{4} - 4 | \Theta | x + 4 \Theta ^{2}}}
                     {- x ^{3} + 2 x - 2 | \Theta |}
            \right) ^{
                      \Sign
                      \left\{
                        \left( 1 - x ^{2} \right)
                        \left( x  - | \Theta | \right) ^{2}
                      \right\}
                     }
        \right\}
        \rmd x
        .
        \nonumber
\end{eqnarray}
The above integral has to be computed when the turning point $x _{\mathrm{min}}$
actually exists, i.e. in the range for $\Theta$ specified by
\eref{eq:theranboutra}. This means that the
$\Sign$ at the exponent is always $+1$, and we can forget about it,
so that the above turns into
\begin{equation}
    \fl
    \bar{S}
    =
    2
    \int _{0} ^{x _{\mathrm{min}}}
        \bar{P} (x) \rmd x
    =
    2
    \int _{0} ^{x _{\mathrm{min}}}
        x
        \tanh ^{-1}
        \left\{
                \frac{x \sqrt{x ^{4} - 4 | \Theta | x + 4 \Theta ^{2}}}
                     {x ^{3} - 2 x + 2 | \Theta |}
        \right\}
        \rmd x
        .
\label{eq:admactint}
\end{equation}
We note that when $x = \Theta$, i.e. the shell is crossing
a (double) zero of the external extremal
Rei\ss{}ner-Nordstr\"{o}m spacetime, the momentum is ill
defined, since the argument of the inverse hyperbolic tangent
is $-1$. $x = \Theta$ is thus a singularity on the
integration path, but, being of the logarithmic type, it is integrable
(the leading contribution to the singularity is determined
in \ref{app:sinintpat}).

\begin{figure}
    \begin{center}
        \fbox{\includegraphics[width=6cm]{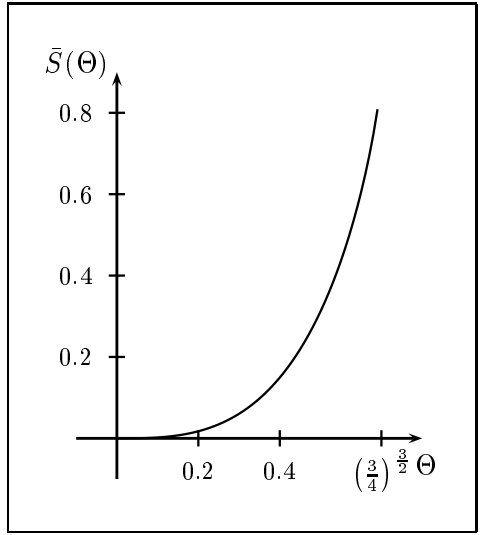}}
    \end{center}
    \caption{Graph of the functional dependence of the action from the
    $\Theta$ parameter as results from the numerical integration of
    \eref{eq:admactint} for $20000$ test values
    of $\Theta$ in the interval $[ 10 ^{-16} , (3/4) ^{3/2}]$ chosen as described
    in the text.}
    \label{fig:actintres}
\end{figure}

The integral in \eref{eq:admactint} is not exactly computable analytically,
but being reassured by the considerations above about its existence,
the integration can be performed numerically. We have performed this kind
of analysis with \texttt{Mathematica}$^{\circledR}$, evaluating
the integral numerically for $10000$ equally spaced test values of $\Theta$ in the
interval $[ 10 ^{-4} , (3/4) ^{3/2}]$ and for other $10000$ test values in the interval
$[ 10 ^{-16} , 10 ^{-4} ]$ taken as a sequence converging to $0 ^{+}$ as $1 / n ^{4}$
for $n \to + \infty$.
The final result is plotted in \fref{fig:actintres}.

\section{\label{sec:WKBquadyn}WKB Quantum States}

We now assume that the system is in a quantum regime; we will perform
its semiclassical quantization \textit{a la} Bohr--Sommerfeld, i.e.
considering the action as an integer multiple of\footnote{We work
in units where $l _{P} ^{2} = \hbar = c = G \equiv 1$.} $\hbar$.
Remebering that all the computations of the previous section are in terms of the
adimensional variables defined in \eref{eq:admdef} and \eref{eq:admPS},
we have
\begin{equation}
    S \left( Q , H \right)
    =
    H ^{2}
    \bar{S} \left( \Theta \right)
    =
    H ^{2}
    \bar{S} \left( \frac{Q}{H} \right)
    ,
    \label{eq:SSbrel}
\end{equation}
so that we can rewrite the Bohr-Sommerfeld quantization condition as
\begin{equation}
    S \left( Q , H \right)
    =
    l _{\mathrm{P}} ^{2} n
    =
    n
    \quad , \quad
    n = 0 , 1 , 2 , \dots
    .
    \label{eq:B-Squacon}
\end{equation}

\subsection{\label{sec:WKBquadynpreext}Preliminary Estimate}

We first note that the quantization can be interpreted as giving a
relation among the parameters of the model, in our case the charge $Q$ and the
de~Sitter cosmological horizon $H$. Let us take for example a total action of the
order of the quantum, $l _{\mathrm{P}} ^{2}$, i.e. associated to a state with
quantum number $n = 1$, for a de~Sitter cosmological horizon $H = \sqrt{2}$.
Then $\bar{S} \approx 1 / 2$, i.e. $\Theta = Q/H \approx 0.55$, so that
$Q \approx 0.78$.
This shows, as a preliminary estimate, that a \textit{small gravitational system}
with quantum properties is conceivable.

\begin{figure}
    \begin{center}
        \fbox{\includegraphics[width=8cm]{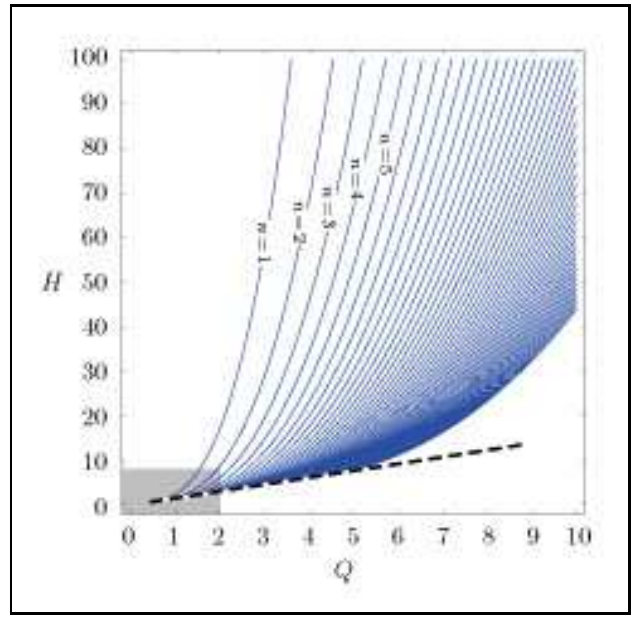}}
    \end{center}
    \caption{Graph of the level curves for the action for the level values
    $1 , 2 , \dots , 50$. The levels are obtained using the numerically
    evaluated action with the function \texttt{ContourPlot} of
    \texttt{Mathematica}$^{\circledR}$. The thicker dashed black line
    displays the limit $H = (4/3) ^{3/2} Q$ of the condition
    $Q / H > (3/4) ^{3/2}$ for which bounded trajectories exist. The grayed
    region is shown, blew up, in {\protect\fref{fig:blowup}}.}
    \label{fig:actlevcur}
\end{figure}

\begin{figure}
    \begin{center}
        \fbox{\includegraphics[width=8cm]{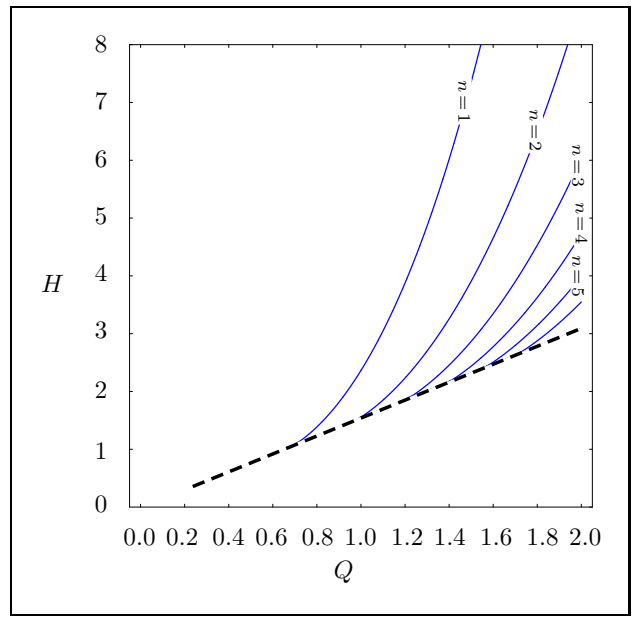}}
    \end{center}
    \caption{Blow up of the region of small $Q$, $H$ parameters.
    This better shows that the quantization condition implies a
    non-vanishing minimum allowed value for both the charge $Q$ and the
    de~Sitter horizon $H$, i.e. a non-vanishing \textit{minimum} allowed value
    for the charge $Q$ together with a \textit{maximum} allowed value for the
    cosmological constant $1/H$.}
    \label{fig:blowup}
\end{figure}

We can also see how the action behaves for variations of the parameters $Q$ and $H$
by a numerical plot of the level curves of the action. This is shown in \fref{fig:actlevcur}.
To get a clearer plot in the region close to the limiting line
$H = (4/3) ^{3/2} Q$ for small $Q$, a smaller region of the plot
is shown, enlarged, in \fref{fig:blowup}.

\subsection{\label{sec:WKBquadynappact}Approximating the action}

\begin{figure}
    \begin{center}
        \fbox{\includegraphics[width=6cm]{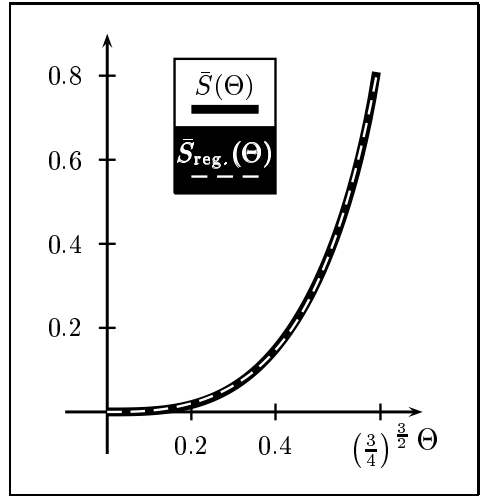}}
    \end{center}
    \caption{Comparison between approximated and numerical evaluated actions.
    The expression $\bar{S} _{\mathrm{reg.}} (\Theta)$ is obtained, as described
    in the text, by a fit to a suitable polynomial. This figure aims to show that
    the main behaviour can be qualitatively approximated, but must not be taken
    to imply that the approximation is everywhere very good. Indeed when the action
    is very small, a very small approximation error can still give a relevant
    relative uncertainty.}
    \label{fig:appnumcom}
\end{figure}

\begin{figure}
    \begin{center}
        \fbox{\includegraphics[width=8cm]{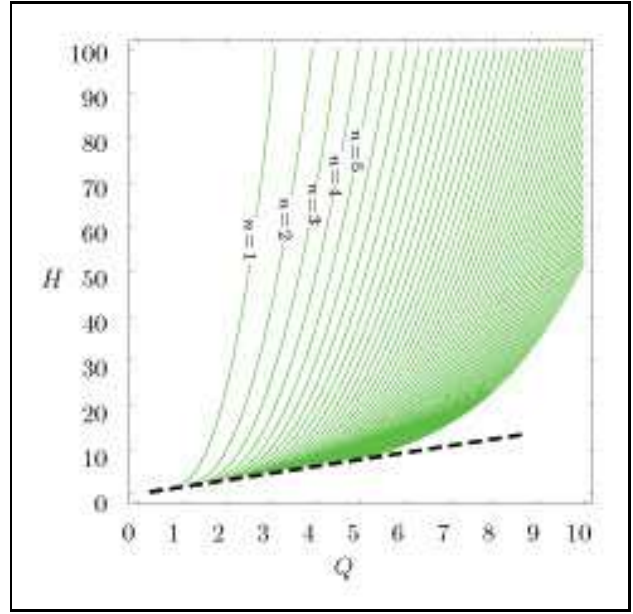}}
    \end{center}
    \caption{Graph of the level curves for the approximated action levels
    $1 , 2 , \dots , 50$. The levels are obtained using the approximating
    polynomial action with the function \texttt{ContourPlot} of
    \texttt{Mathematica}$^{\circledR}$. The thicker dashed black line
    displays the limit $H = (4/3) ^{3/2} Q$ of the condition
    $Q / H > (3/4) ^{3/2}$ for which bounded trajectories exist.}
    \label{fig:reglevcur}
\end{figure}

To get some analytical result we try to fit the $20000$ points for which
we evaluated the action with some simple (polynomial) function. In this way we
will be able to get an (approximated) relation among $Q$, $H$ and $n$.
The choice of the approximating function is done with two goals in mind:
\begin{enumerate}
    \item to approximate in the best possible way the behaviour
    of the action $\bar{S}$ at least in some regime;
    \item to have a simple enough expression to get an algebraic
    relation among $Q$, $H$ and $n$.
\end{enumerate}
To fulfill the above requirements we first analyze the leading $\Theta$
dependence of the action $\bar{S}$, expressed as the integral
\eref{eq:admactint}. From \ref{app:powexp} we se that
$\bar{P} (x) \sim x ^{2} + \Or (x) ^{2}$
and the upper turning point
$x _{\mathrm{min}} (\Theta) \sim \Theta + \Or (\Theta)$,
so that
$\bar{S} (\Theta) \sim \Theta ^{3} + \Or (\Theta) ^{3}$.
We thus choose an approximating function starting with a third power of
$\Theta$ and having the next two powers of $\Theta$:
\begin{equation}
    A x ^{5} + B x ^{4} + C x ^{3}
    ,
\label{eq:apppol}
\end{equation}
The coefficients $A$, $B$, $C$ in \eref{eq:apppol} have been determined from the
$20000$ sample points using the function \texttt{Regress} of
\texttt{Mathematica}$^{\circledR}$, with the parameter
\texttt{IncludeConstant} $\to$ \texttt{False}, since there is no constant
term in the model:
they result to be, together with the corresponding standard errors,
\begin{eqnarray}
    A = + 10.11 \pm 0.04
    \nonumber \\
    B = - 7.39 \pm 0.04
    \label{eq:admfitcoe}
    \\
    C = + 3.64 \pm 0.01
    \nonumber
    ,
\end{eqnarray}
with an \textit{adjusted regression coefficient} of $0.99981$.
\\
The comparison between approximated and numerical evaluated actions
is plotted in \fref{fig:appnumcom}.
Level curves of the approximated action are plotted in \fref{fig:reglevcur}.

From the approximated expression \eref{eq:apppol}, which
by \eref{eq:SSbrel} and \eref{eq:B-Squacon} must equal $n$ when
multiplied by $H ^{2}$, we can get the following equation
of the third degree in H,
\begin{equation}
    A Q ^{5}
    +
    B Q ^{4} H
    +
    C Q ^{3} H ^{2}
    -
    n H ^{3}
    =
    0
    :
\label{eq:QHnpolequ}
\end{equation}
this can be solved exactly to get the approximated
relation for $H$ as a function of $Q$ and $n$ as it comes
from the simplified expression \eref{eq:apppol} for the action.
The quite complex algebraic expression is
\begin{eqnarray}
    \fl
    H (Q ; n)
    =
    \frac{C Q ^{3}}{3 n}
    +
    \frac{2 ^{1/3} H _{1} (Q ; n) Q ^{7/3}}
         {3 n \sqrt[3]{H _{2} (Q ; n) + \sqrt{4 Q ^{2} H _{1} ^{3} (Q ; n) + H _{2} ^{2} (Q ; n)}}}
    \nonumber \\
    -
    \frac{Q ^{5/3}\sqrt[3]{H _{2} (Q ; n) + \sqrt{4 Q ^{2} H _{1} ^{3} (Q ; n) + H _{2} ^{2} (Q ; n)}}}{2 ^{1/3} 3 n}
    ,
    \label{eq:HQnrel}
\end{eqnarray}
where
\begin{eqnarray}
    \fl
    H _{1} (Q ; n)
    =
    - C ^{2} Q ^{2} - 3 B _{1} n
    \nonumber \\
    \fl
    H _{2} (Q ; n)
    =
    - 2 C ^{3} Q ^{4} - 9 B C n Q ^{2} - 27 A ^{2} n ^{2}
    \nonumber
\end{eqnarray}
and the approximated relations above are only valid if
$$
    H > \frac{| Q |}{(3/4) ^{3/2}}
    ,
$$
a condition plainly coming from \eref{eq:theranboutra}.
The approximated levels of \fref{fig:reglevcur} are nothing
but the graphs of the $n = \mathrm{const.}$ relations coming from \eref{eq:HQnrel}.

\section{\label{sec:dis}Discussion}

We have presented a model in which it is analyzed
a general relativistic system composed of two spacetime domains, a
de~Sitter interior with cosmological horizon $H$
and an extremal Rei\ss{}ner-Nordstr\"{o}m exterior with
$M = | Q |$,
joined across a thin shell of mass-energy $m = | Q |$. A semiclassical
quantization of classically bounded trajectories can be performed using
a scheme \textit{a la} Bohr--Sommerfeld. In this way it is found that
the quantum states can be characterized by a quantum number $n$, which
is responsible for restricting the allowed values of $Q$ and $H$. In
particular the quantum dynamics of the system constrains the values
of $H$ as functions $H = H (Q ; n)$, which we have approximated after
a numerical analysis of the problem.

It is interesting to note that this is a sensible result in a semiclassical
approximation to a quantum gravitational situations. Indeed a full quantum theory
of gravity would treat the three-metric as a dynamical infinite-dimensional
variable to be determined, say, by the Wheeler--deWitt equation in superspace.
From our point of view, under the assumption we made, we are only in a minisuperspace
approximation, since the functional form of the metric functions
is fixed from the very beginning, leaving $H$ and $Q$ as the only free parameters:
quantum gravity will thus impose condition on these quantities, i.e. determine
the only residual degrees of freedom in the three geometry.
We can also see from the enlarged plot of \fref{fig:blowup}, that the
quantization condition selects a \textit{minimum} value
for the allowed charge of the semiclassical quantized shell. At the same time,
the cosmological constant of the interior region, cannot exceed a
\textit{maximum} value, which is also a sensible consequence of quantization.
For a given fixed value of the ratio $Q/H = \bar{\kappa}$, which means that
we are ``moving'' across the graphs of figures \ref{fig:actlevcur} and
\ref{fig:blowup} along a line through the origin, we have discrete allowed
values for $Q$ and $H$: these point are located on a parabola,
since from \eref{eq:SSbrel} we exactly have
$S (\bar{\kappa} H , H ) = H ^{2} \bar{S} (\bar{\kappa})$.

We also note that by the final result,
systems characterized by different scales in the parameters can be described:
among these, as expected, we find small scale systems
(with a charge which is a small multiple of the elementary electron charge).
Moreover an external asymptotic observer measures
a total mass energy for the shell given by
\[
    E = m + \frac{1}{2} \left( \frac{Q ^{2}}{r ^{2}} - \frac{M ^{2}}{r ^{2}} \right)
\]
which in our case is nothing but $E = m$, since $M ^{2} = Q ^{2}$: thus we effectively
see in the outside domain an object with rest mass $m = | Q |$ and charge $Q$,
which is the exterior manifestation of a bounded interior containing a part
of spacetime characterized by a non-vanishing vacuum energy. Due to the form of
the potential this bound semiclassical state is metastable, and will decay into
an infinitely expanding shell after a finite time
(a similar situation occurs in \cite{1999ClQuGr.16..3315..G}),
which in principle could be calculated (as proper time), studying the process
of tunnelling across the classical effective potential barrier\footnote{This
will be the topic of a forthcoming paper.}.

\appendix

\section{\label{app:potcritic}Zeros of the potential and critical value of $\Theta$}

The zeros of the potential $\bar{V} (x)$ can be obtained in closed form, since they are
the zeroes of the numerator, i.e. the solutions of the equation
\begin{equation}
    \bar{V} (x) = 0
    \quad \Rightarrow \quad
    x ^{2} (x ^{4} - 4 a x + 4 a ^{2}) = 0
\label{eq:potzerequapp}
\end{equation}
with $a > 0$ and $- x ^{3} + 2 x - 2 a \neq 0$.
We are interested in the non-negative solutions, which, apart from the
$x = 0$ one, can be determined exactly as solutions of the fourth order equation
\begin{equation}
    x ^{4} - 4 a x + 4 a ^{2} = 0
    .
\label{eq:potzer}
\end{equation}
By setting
\[
    {\mathcal{B}}
    =
    \left\{
        9 a ^{2} + \left( 81 a ^{4} - 192 a ^{6} \right) ^{1/2}
    \right\} ^{1/3}
\]
we have that for $0 < a \leq (3/4) ^{3/2}$
\begin{equation}
    \fl
    x _{\stackrel{\mathrm{\scriptstyle{}min}}{\mathrm{\scriptstyle{}max}}}
    =
    \frac{1}{3 ^{1/3} 2 ^{1/2}}
    \left\{
        \sqrt{\frac{4 \cdot 3 ^{1/3} a ^{2}}{{\mathcal{B}}} + {\mathcal{B}}}
        \mp
        \sqrt{
            -
            \frac{4 \cdot 3 ^{1/3} a ^{2}}{{\mathcal{B}}}
            -
            {{\mathcal{B}}}
            +
            \frac{6 \cdot 2 ^{1/2} a}
                 {\sqrt{\frac{4 \cdot 3 ^{1/3} a ^{2}}{{\mathcal{B}}} + {\mathcal{B}}}}
        }
    \right\}
    .
\label{eq:xmixma}
\end{equation}
These expressions are not very enlightening: the one for $x _{\mathrm{min}}$ has
been used to exactly evaluate the upper integration limit in the numerical evaluation of
the integral that gives the classical action.

To see when the quartic part of the potential has two positive roots we can use the expression
above, but also a smarter procedure, as follows.
Clearly the limiting case is the one in which the potential is tangent to the
positive $x$ axis, i.e. the two solutions coincide. In this case the quartic
part must be of the form
\begin{eqnarray}
    \fl
    (x - \alpha) ^{2} (x ^{2} + \beta x + \gamma)
    \nonumber \\
    =
    x ^{4}
    +
    ( \beta - 2 \alpha ) x ^{3}
    +
    ( \gamma - 2 \alpha \beta + \alpha ^{2}) x ^{2}
    +
    ( \alpha \beta ^{2} - 2 \alpha \gamma) x
    +
    \alpha ^{2} \gamma
    \nonumber
    ,
\end{eqnarray}
which by comparison with $x ^{4} - 4 a x - 4 a ^{2}$ gives the set of equations
\[
    \left \{
    \matrix{
        \beta - 2 \alpha = 0
        \hfill
        \cr
        \gamma - 2 \alpha \beta + \alpha ^{2} = 0
        \hfill
        \cr
        \alpha ^{2} \beta - 2 \alpha \gamma = - 4 a
        \hfill
        \cr
        \alpha ^{2} \gamma = 4 a ^{2}
        \hfill
        }
    \right .
    .
\]
Then $\alpha = \beta / 2$ and $\gamma = 3 \beta ^{2} / 4$ from the first two equations.
This gives $\alpha = a ^{1/3}$ from the third and
$\alpha = (4 a ^{2} / 3) ^{1/4}$ from the fourth. These last two relations are
compatible for non-vanishing $a$, if and only if $a = (3/4) ^{3/2}$, which is
thus the critical value of $a$.

\section{\label{app:staiss}Stability of the shell against single particle decay}

In this section we discuss the stability of the trajectory of the infinitesimally
thin shell under single particle decay, following the treatment
that can be found in \cite{1970PhReLe.25..1771..G}.
The proof that the shell is stable against single particle decay of uncharged particles
is not reproduced here, because it can be easily derived from the reference cited
above, to which the reader is referred. We will instead shortly discuss the case in which
charged particles are involved.

In more detail, the relevant question is if the motion of a charged particle,
which at the instant of maximum expansion starts out where the shell is located,
will subsequently be governed by a confining, i.e. ``$\cup$-shaped'',
effective potential, or not.
Performing the analysis at the instant of maximum expansion, where the potential
is static, simplifies the computation: subsequent changes of the potential will
have, anyway, only adiabatic effects on the locally trapped particle and this is
not relevant for the point under discussion.
To get the desired result we will proceed in two steps:
\begin{enumerate}
    \item identify the effective potential governing the motion of a particle
    in the exterior Rei\ss{}ner-Nordstr\"{o}m geometry;
    \item evaluate if it is ``$\cup$-'' or ``$\cap$-shaped'' at the point of maximum
    expansion.
\end{enumerate}
We will work in the adimensional units used throughout the rest of the paper.

\subsection{Effective potential for a charged particle in the Rei\ss{}ner-Nordstr\"{o}m
spacetime}

The effective potential for the motion of a particle of charge $q$ and mass $\mu$ in
the Rei\ss{}ner-Nordstr\"{o}m spacetime can be obtained in many different ways. Probably
the quickest one is to start from the more general result that can be found in the
equation after equation (3) in Box~33.5 of \cite{1970WHFranCGr........M}, i.e. the effective
potential for the orbits of test particles in the equatorial plane of a Kerr-Newman
black hole. Specializing this result to a black hole with zero
angular momentum we obtain the effective potential in the Rei\ss{}ner-Nordstr\"{o}m case.
Perhaps more instructive is to
perform again the analysis until equation (6) of \cite{1970PhReLe.25..1771..G}
adding the electrostatic contribution to the particle momentum or solving the
Hamilton-Jacobi equation in the Rei\ss{}ner-Nordstr\"{o}m metric. Anyway, the final
result for the extremal case we are interested in is
\begin{equation}
    \tilde{V} (x)
    =
    \frac{\tilde{\epsilon} \Theta}{x}
    +
    \left( 1 - \frac{\Theta}{x} \right)
    \left( 1 + \frac{\tilde{\lambda} ^{2}}{x ^{2}} \right) ^{1/2}
    ,
\label{eq:effreinorpot}
\end{equation}
where $\tilde{\epsilon}$ is the charge/mass ratio of the test particle and
$\tilde{\lambda} = \tilde{L} / H$ is its angular momentum per unit mass
($\tilde{L} = L / \mu)$ in the $H$ scale defined in \eref{eq:admdef} and \eref{eq:admPS}.

\subsection{Evaluation of the Effective Potential at the point of maximum expansion}

With the above results at hand, we now evaluate the second derivative of the
effective potential, i.e.
\begin{eqnarray}
    \fl
    \tilde{V} (x) ''
    =
    \frac{
        3 x ^{2} \tilde{\lambda} ^{2} +
        2 \tilde{\lambda} ^{4} +
        2 \tilde{\epsilon} \Theta
        \left( x ^{2} + \tilde{\lambda ^{2}} \right) ^{3/2}
        -
        (2 x ^{4} + 9 x ^{2} \tilde{\lambda} ^{2} + 6 \tilde{\lambda} ^{4})
        | \Theta |
         }
         {x ^{3} \left( x ^{2} + \tilde{\lambda ^{2}} \right) ^{3/2}}
    ;
\label{eq:effreinorpotsecder}
\end{eqnarray}
we are interested in its sign at the value $x _{\mathrm{min}}$,
the maximum radius of the shell given in \eref{eq:xmixma}: a positive
sign will indicate that the potential is ``$\cup$-shaped'' and thus
the shell stable. To get the result, we can restrict the study to the
numerator $\tilde{N} (x _{\mathrm{min}})$ of \eref{eq:effreinorpotsecder},
$$
    \tilde{N} (x _{\mathrm{min}})
    =
    3 x _{\mathrm{min}} ^{2} \tilde{\lambda} ^{2} +
    2 \tilde{\lambda} ^{4} +
    2 \tilde{\epsilon} \Theta
    \left( x ^{2} _{\mathrm{min}} + \tilde{\lambda ^{2}} \right) ^{3/2}
    -
    (2 x _{\mathrm{min}} ^{4} + 9 x _{\mathrm{min}} ^{2} \tilde{\lambda} ^{2} + 6 \tilde{\lambda} ^{4})
    | \Theta |
    ,
$$
since the denominator does not contribute to the sign.
We also remember that the quantity $x _{\mathrm{min}}$ depends on $\Theta$, since
\eref{eq:potzerequapp} comes from \eref{eq:admpot} with $a = | \Theta |$.

Even with the above simplification, the detailed study of the sign of the quantity under
consideration is complicated, mainly because of the non-trivial $\Theta$ dependence;
a \textit{graphical} analysis is also of little help, since $\tilde{N} (x _{\mathrm{min}})$
depends on the three variables $\Theta$, $\tilde{\epsilon}$ and $\tilde{\lambda}$,
namely the adimensional charge $Q/H$, the charge/mass ratio $q / \mu$ of the test particle
and the adimensional angular momentum per unit mass $L / (H \mu)$ of the test particle.
Thus we will \textit{not} search for the most general result, i.e. we will not give
\textit{necessary and sufficient} conditions for the stability of the shell;
we will show, instead, that in some physically reasonable situations the shell itself
is indeed stable against single charged particle decay.

In particular we see that in the units we are using, the adimensional parameter
$\tilde{\epsilon}$ is a large number, i.e. the charge/mass ratio for an elementary
particle is very large. Thus we consider the behaviour of the numerator for large
$\tilde{\epsilon}$:
$$
    \lim _{\tilde{\epsilon} \to \infty}
    \tilde{N} (x _{\mathrm{min}})
    =
    + \infty
    ;
$$
the limit has always the ``$+$'' sign, since we consider emission of charges of the
same sign of those composing the shell; thus in this limit the second derivative
of the effective potential is positive, which shows stability under charged elementary
particle decay.

As a second case, we see what happens for radial emission of particles:
$$
    \lim _{\tilde{\lambda} \to 0}
    \tilde{N} (x _{\mathrm{min}})
    =
    2 x _{\mathrm{min}} ^{3} | \Theta | ( | \tilde{\epsilon} | - x _{\mathrm{min}})
    ;
$$
we again used the fact that ejected particles have the same charge as the shell, so that
$\tilde{\epsilon} \Theta = |\tilde{\epsilon}| \cdot |\Theta|$.
In this case also we see that, for elementary particle emission, we certainly can realize the situation
$\tilde{\epsilon} > x _{\mathrm{min}}$, so the sign is again positive.

Even restricting the study to the two cases above, we can thus conclude that, shell configurations
which are stable against single particle decay \textit{can be realized}.

\section{\label{app:sigintsigout}General analysis for $\sigma _{\mathrm{in}}$,
$\sigma _{\mathrm{out}}$ on a bounded trajectory}

\begin{figure}
    \begin{center}
        \fbox{\includegraphics[width=6cm]{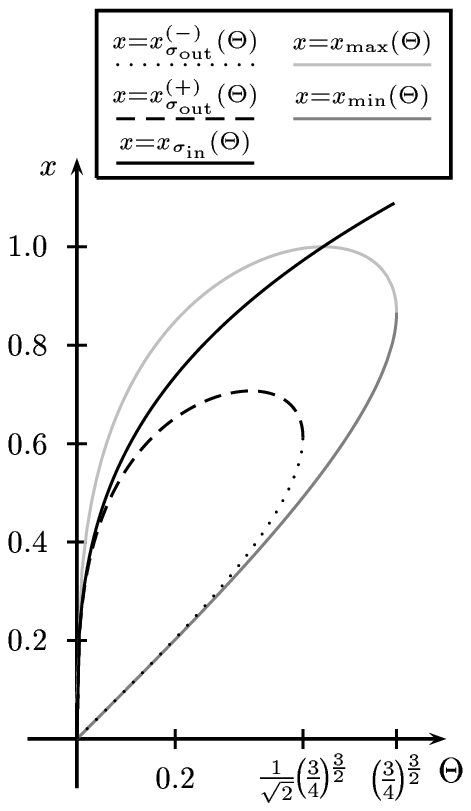}}
    \end{center}
    \caption{Graph of the curves that helps in determining the sign
    of $\sigma$'s for different values of the parameter $\Theta$. Since
    the classically allowed trajectories are delimited by the
    $x = 0$ axis and by the dark gray continuous curve
    $x = x _{\mathrm{min.}} (\Theta)$, we see that no changes of
    signs of $sigma$'s occur along them.}
    \label{fig:sigsig}
\end{figure}

The ``graphical'' analysis of the classical dynamics performed in
\sref{sec:cladyn} is based on the plot of \fref{fig:clastu}.
In the discussion a relevant point is the sign of
$\bar{\sigma} _{\mathrm{in}}$ and $\bar{\sigma} _{\mathrm{out}}$, which is crucial in
determining the direction of the normal to the shell pointing in
the direction of increasing radius as well as the side of the Penrose diagram
crossed by the trajectory. We will show here that the situation analyzed for
the particular value $\Theta = 0.55$ is, in fact, general. The sign of
$\bar{\sigma} _{\mathrm{in}}$ is given \eref{eq:admsigin}, so that we see
that is positive for
\[
    x < x _{\sigma _{\mathrm{in}}} ( \Theta ) \equiv ( 2 \Theta ) ^{1/3}
    .
\]
More complicate is the analysis of the sign of $\bar{\sigma} _{\mathrm{out}}$, because
from \eref{eq:admsigout} we see that it is determined as the sign of a polynomial
of order four. It has at most two real roots for
$\Theta < \Theta _{\mathrm{crit.}} / \sqrt{2} = (3/4) ^{3/2} / \sqrt{2}$; this
can be seen also from the analysis of \ref{app:potcritic}. Indeed
$\bar{\sigma} _{out} > 0$ if
\[
    x ^{4} - 2 a x + 2 a ^{2} < 0 \quad \mathrm{with} \quad a > 0
    .
\]
The real roots of the left hand side of the above inequality can
be deduced from those of the left hand side of \eref{eq:potzer}
since
\[
    \fl
    x ^{4} - 2 a x + 2 a ^{2}
    \longrightarrow
    \frac{1}{4}
    \left(
        x ^{4} - 4 a x + 4 a ^{2}
    \right)
    \quad
    {\mathrm{when}}
    \quad
    x \to \frac{x}{\sqrt{2}}
    \quad \mathrm{and} \quad
    a \to \frac{a}{\sqrt{2}}
    .
\]
Then, if we call $x _{\sigma _{\mathrm{out}}} ^{(-)}$ and
$x _{\sigma _{\mathrm{out}}} ^{(+)}$
the two real values for which $\bar{\sigma} _{\mathrm{out}}$ changes sign,
we see, using the transformation above, that we must have
$a \leq (3/4) ^{3/2} / \sqrt{2}$
and that
\begin{equation}
    \fl
    x _{\sigma _{\mathrm{out}}} ^{(\mp)}
    =
    \frac{1}{2 \cdot 3 ^{1/3}}
    \left\{
        \sqrt{\frac{8 \cdot 3 ^{1/3} b ^{2}}{{\mathcal{T}}} + {\mathcal{T}}}
        \mp
        \sqrt{
            -
            \frac{8 \cdot {3 ^{1/3}}  b^{2}}{{\mathcal{T}}}
            -
            {\mathcal{T}}
            +
            \frac{12 b}{\sqrt{\frac{8 \cdot  3 ^{1/3} b ^{2}}{{\mathcal{T}}} + {\mathcal{T}}}}
             }
    \right\}
\label{eq:xsmixsma}
\end{equation}
with
\[
    {\mathcal{T}}
    =
    \sqrt[3]{18 b ^{2} + 2 \sqrt{81  b ^{4} - 384 b ^{6}}}
\]
Since the above expressions are not very enlightening (and the same is true
for those of \eref{eq:xmixma}), the easiest way to compare them with
the turning points is again the graphical one, i.e. the plot of
$x _{\sigma _{\mathrm{in}}} (\Theta)$,
$x _{\sigma _{\mathrm{out}}} ^{(\pm)} (\Theta)$,
$x _{\mathrm{min}} (\Theta)$,
$x _{\mathrm{max}} (\Theta)$,
which we can see in \fref{fig:sigsig}.
For small $x$ (on the vertical axis) $\bar{\sigma } _{\mathrm{in}}$ is positive
and $\bar{\sigma} _{\mathrm{out}}$ is negative. Since all the zeroes, when they
exist, are bigger than the the turning point $x _{\mathrm{min}}$, which is the
upper limit of the bounded trajectory, then there is no change of
sign of $\sigma$'s along it.

\section{\label{app:sinintpat}Character of the singularity on the integration path}

In this section we determine the leading contribution to the logarithmic
(and thus integrable) singularity on the integration path that appears
in the evaluation of the integral in \eref{eq:admactint}. The logarithmic
character stems from the definition of the inverse hyperbolic tangent in terms
of the logarithm and from the fact that its argument is a rational function
with the following properties\footnote{We define ${\mathcal{F}}$, ${\mathcal{R}}$
and ${\mathcal{D}}$ according to the first two $\equiv$'s of the equation
below.}:
\[
    \lim _{x \to | \Theta |}
        {\mathcal{F} (x ; \Theta)}
    \equiv
    \lim _{x \to | \Theta |}
        \frac{{\mathcal{R}}  ^{1/2} (x ; \Theta)}{{\mathcal{D}} (x ; \Theta)}
    \equiv
    \lim _{x \to | \Theta |}
        \frac{x \sqrt{x ^{4} - 4 | \Theta | x + 4 \Theta ^{2}}}{ x ^{3} - 2 x + 2 | \Theta |}
    =
    1
    ,
\]
i.e.
\begin{equation}
    {\mathcal{F}} ( | \Theta | ; \Theta ) = 1
\label{eq:Fn0}
\end{equation}
To extract the behaviour of the above function around the point $x = | \Theta |$
we develop it in power series. Let us set
\begin{eqnarray}
    {\mathcal{R}} (x ; \Theta)
    =
    x ^{6} - 4 | \Theta | x ^{3} + 4 \Theta ^{2} x ^{2}
    \quad
    \mathrm{with}
    \quad
    {\mathcal{R}} (| \Theta | ; \Theta) = \Theta ^{6}
    \\
    {\mathcal{D}} (x ; \Theta)
    =
    x ^{3} - 2 x + 2 | \Theta |
    \quad
    \mathrm{with}
    \quad
    {\mathcal{D}} (| \Theta | ; \Theta) = | \Theta | ^{3}
    .
\end{eqnarray}
Then it follows
\begin{eqnarray}
    \fl
    {\mathcal{R}} ' (x ; \Theta)
    =
    6 x ^{5} - 12 | \Theta | x ^{2} + 8 \Theta ^{2} x
    \quad
    \mathrm{so\ that}
    \quad
    {\mathcal{R}} ' (| \Theta | ; \Theta) = 6 | \Theta | ^{5} - 4 | \Theta | ^{3}
    \\
    \fl
    {\mathcal{D}} ' (x ; \Theta)
    =
    3 x ^{2} - 2
    \quad
    \mathrm{so\ that}
    \quad
    {\mathcal{D}} ' (| \Theta | ; \Theta) = 3 \Theta ^{2} - 2
    .
\end{eqnarray}
and
\begin{eqnarray}
    \fl
    {\mathcal{R}} '' (x ; \Theta)
    =
    30 x ^{4} - 24 | \Theta | x + 8 \Theta ^{2}
    \quad
    \mathrm{so\ that}
    \quad
    {\mathcal{R}} '' (| \Theta | ; \Theta) = 30 \Theta ^{4} - 16 \Theta ^{2}
    \\
    \fl
    {\mathcal{D}} '' (x ; \Theta)
    =
    6 x
    \quad
    \mathrm{so\ that}
    \quad
    {\mathcal{D}} '' (| \Theta | ; \Theta) = 6 | \Theta |
    .
\end{eqnarray}
We then compute
${\mathcal{F}} ' (| \Theta | ; \Theta)$ and ${\mathcal{F}} '' (| \Theta | ; \Theta)$,
i.e. the first and second derivatives of the argument of the inverse hyperbolic tangent.
Since
\[
    {\mathcal{F}} '
    =
    \frac{{\mathcal{R}} ' {\mathcal{D}} - 2 {\mathcal{R}} {\mathcal{D}} '}
         {2 {\mathcal{R}} ^{1/2} {\mathcal{D}} ^{2}}
\]
we obtain
\begin{equation}
    {\mathcal{F}} ' ( | \Theta | ; \Theta ) = 0
    .
\label{eq:Fn1}
\end{equation}
Moreover
\[
    {\mathcal{F}} ''
    =
    \frac{
          2
          \left(
                {\mathcal{R}} '' {\mathcal{D}}
                -
                {\mathcal{R}} ' {\mathcal{D}} '
                -
                2 {\mathcal{R}} {\mathcal{D}} ''
          \right)
          {\mathcal{R}} {\mathcal{D}}
          -
          \left(
                {\mathcal{R}} ' {\mathcal{D}}
                -
                2 {\mathcal{R}} {\mathcal{D}} '
          \right)
          \left(
                {\mathcal{R}} ' {\mathcal{D}}
                +
                4 {\mathcal{R}} {\mathcal{D}} '
          \right)
         }
         {4 {\mathcal{R}} ^{3/2} {\mathcal{D}} ^{3}}
\]
and we get
\begin{equation}
    {\mathcal{F}} '' ( | \Theta | ; \Theta)
    =
    - \frac{4}{\Theta ^{6}} \left( 1 - \Theta ^{2} \right)
    .
\label{eq:Fn2}
\end{equation}
The expansion of ${\mathcal{F}}$ around $x = | \Theta |$ up to second order can
then be written, using \eref{eq:Fn0}, \eref{eq:Fn1} and \eref{eq:Fn2}, as
\[
    {\mathcal{F}} (x ; \Theta)
    =
    1
    -
    \frac{2 \left( 1- \Theta ^{2} \right)}{\Theta ^{6}}
    \left( x - | \Theta | \right) ^{2}
    +
    \Or \left( x - | \Theta | \right) ^{3}
\]
and inserting this result inside the expression of the hyperbolic
tangent in terms of logarithms, we can easily see, as expected, that the
singularity is integrable.

\section{\label{app:powexp}$\Theta$ dependence of the Action}

We consider the action integral \eref{eq:admactint}. Expanding the integrand
we get
\[
    x
    \tanh ^{-1}
    \left\{
            \frac{x \sqrt{x ^{4} - 4 | \Theta | x + 4 \Theta ^{2}}}
                 {x ^{3}- 2 x + 2 | \Theta |}
    \right\}
    =
    x ^{2}
    +
    \frac{1}{2 | \Theta |} x ^{3}
    +
    \Or (x ^{4})
    ;
\]
Then the upper integration limit, $x (\Theta)$, can be expanded as
\[
    x ( \Theta )
    =
    \Theta
    +
    \Or ( \Theta ^{2})
\]
so that when both expansions hold, we can write
\[
    \bar{S} (\Theta)
    \sim
    \frac{11 | \Theta | ^{3}}{12}
    +
    \Or (\Theta ^{4})
    .
\]
We thus see that the leading term for small $| \Theta |$ is $\sim | \Theta ^{3} |$.

\ack

The author wish to thank the Department of Physics and Astronomy of the University of
Victoria for hospitality and Professor Werner Israel for useful discussions, comments
and suggestions relevant to the topics addressed in the paper.

\end{document}